\documentclass[prb,aps,showpacs,amssymb,twocolumn]{revtex4}
\usepackage{epsfig}

\newcommand{\bra}[1]{\langle#1|}

\newcommand{\ket}[1]{|#1\rangle}

\newcommand{\spupup}{\ket{\! \uparrow\uparrow}}

\newcommand{\spupdown}{\ket{\! \uparrow\downarrow}}

\newcommand{\spdownup}{\ket{\! \downarrow\uparrow}}

\newcommand{\spdowndown}{\ket{\! \downarrow\downarrow}}

\begin{document}

\title{Entanglement transfer from electron spins to photons in spin light-emitting diodes containing quantum dots}

\author{Veronica Cerletti, Oliver Gywat, and Daniel Loss}

%\email{veronica.cerletti@unibas.ch}

\affiliation{Department of Physics and Astronomy, University of Basel, Klingelbergstrasse 82, CH-4056 Basel, Switzerland}

%\twocolumn[\hsize\textwidth\columnwidth\hsize\csname
%@twocolumnfalse\endcsname

\begin{abstract}

We show that electron recombination using positively charged excitons in
single quantum dots provides an efficient method to transfer entanglement from
electron spins onto photon polarizations.
We propose a scheme for the production of entangled four-photon states
of GHZ type.
From the GHZ state, two fully entangled photons can be obtained by a
measurement of two photons in the linear polarization basis,
even for quantum dots with observable fine structure splitting for
neutral excitons and
significant exciton spin decoherence.
Because of the interplay of quantum mechanical selection rules and
interference, maximally entangled electron pairs are converted into
maximally entangled photon pairs with unity fidelity for a continuous set
of observation directions.
We describe the dynamics of the conversion process using a
master-equation approach and show that the implementation of our scheme
is feasible with current experimental techniques.
\end{abstract}


% 78.67.-Hc Optical properties of quantum dots
% 71.35.Pq Charged excitons (trions)
% 73.40.-c Electronic transport in interface structures
% 03.67.Mn Entanglement production, characterization, and manipulation

\maketitle

\section{Introduction \label{sec:Intro}}
Spin light-emitting diodes
(spin-LEDs),~\cite{fiederling:1999a,ohno:1999a,awschalom:02,pryor:03,guendogdu:04,seufert,book:02,kroutvar:2004a}
in which electron recombination is accompanied by the emission of a photon with well-defined circular
polarization, provide an efficient interface between electron spins and photons. The operation of such devices
at the single-photon level would allow one to convert the quantum state of an electron encoded in its spin state
into that of a photon with a wide range of possible applications. In view of quantum information schemes,
converting spin into photon quantum states corresponds to a conversion of localized into flying qubits, which
can be transmitted over long distances and could overcome limitations caused by the short-range nature of the
electron exchange interaction.~\cite{book:02} On a more fundamental level, the photon polarization can be
readily measured experimentally such that an interface between spins and photons will allow one to measure
quantum properties of the spin system via the photons generated on recombination. More specifically,
entanglement of electron spins could be demonstrated not only in current noise~\cite{burkard:2000,egues:2002}
but also via photon polarizations which allows one to test Bell's inequalities.~\cite{bell:65}

\begin{figure}[t]
\centerline{\includegraphics[width=7.5cm]{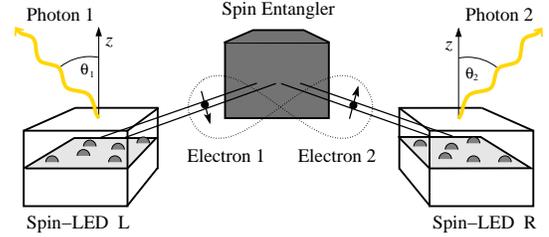}
%\vspace{1mm}
} \caption{(Color online) Schematic setup for the transfer of entanglement between electrons and photons. An
electron entangler (gray box) injects a pair of spin-entangled electrons into two current leads. The electrons
recombine individually in one quantum dot located in the left (L) and one in the right (R) spin-LED and give
rise to the emission of two photons.} \label{fig:setup}
\end{figure}

In this work, we show that nonlocal spin-entangled electron pairs that recombine in single quantum dots
contained in spatially separated spin-LEDs are converted into polarization-entangled photon states. In addition
to its applications in quantum communication, this transfer can be used to characterize the output of an
electron spin
entangler~\cite{andreev:01,lesovik:01,recher:02,bena:02,bouchiat:03,saraga:03,recher:03,saraga2:04} in a setup
as shown in Fig.~\ref{fig:setup}. Furthermore, such a setup acts as a deterministic source of
polarization-entangled photon pairs.
Recently, the decay of biexcitons in single quantum dots has been proposed
for the production of entangled photons.~\cite{benson:00,moreau:01} However, several
experiments~\cite{kiraz:02,santori:02,stevenson:2002a,zwiller:02,ulrich:2003a} have only shown polarization
correlation but not entanglement of the photons. The fine structure splitting $\delta_{\mathrm{ehx}}$ of the
bright exciton ground state~\cite{takagahara:00}  has been identified to be crucial for the lack of
entanglement: Firstly, the polarization-entangled photons are also entangled in energy if
 $\delta_{\mathrm{ehx}}$ is larger than the exciton
linewidth.~\cite{stace:2003a} Secondly, for  $\delta_{\mathrm{ehx}}\neq 0$ the exciton spin relaxation rate due
to phonons $1/T_{1,X}$ is enhanced~\cite{tsitsishvili:2003a} and leads to an increased decoherence rate
$1/T_{2,X} = 1/2T_{1,X} + 1/T_{\varphi,X}$, where $1/T_{\varphi,X}$ is the pure decoherence rate. To overcome
these difficulties we propose to use positively charged excitons ($X^+$), for which $\delta_{\mathrm{ehx}}= 0$
up to small corrections. Moreover, we demonstrate that the antisymmetric hole ground state of the $X^+$ enables
the production of entangled four-photon states. We study the transfer of entanglement for different photon
emission directions by calculating the von Neumann entropy. Due to quantum mechanical interference, the fidelity
of this process approaches unity not only for photon emission along the spin quantization axis, but for a
continuous set of observation directions. The relaxation and decoherence of the electron spins in the leads is
modeled using a master equation and it is quantified by the fidelity of the entangled state.

This work is organized as follows. In Sec.~\ref{sec:Dynamics} we describe the
dynamics of the conversion process. In Sec.~\ref{sec:Optic} we focus on the
microscopic expressions for the involved optical transitions, leading to entangled
four-photon and two-photon states. In Sec.~\ref{sec:Entanglement} we quantify the
entanglement of the two-photon state as a function of the emission angles.
We conclude in Sec.~\ref{sec:Concl}.

\section{Dynamics of the conversion process \label{sec:Dynamics}}
The effective Hamiltonian of the system is given by
%$H=H_{L}+H_{R}+H_{\mathrm{rad}}+H_{\mathrm{int}}$,
\begin{equation}
H=H_{L}+H_{R}+H_{\mathrm{rad}}+H_{\mathrm{int}},
\end{equation}
where $H_{\mathrm{\alpha}}=\mathbf{p}^{2}/2m+V_{\mathrm{qd}}(\mathbf{r})$ is the Hamiltonian of the quantum dot
$\alpha=L,R$ with confinement potential $V_{\mathrm{qd}}(\mathbf{r})$. The Hamiltonian of the radiation field is
$H_{\mathrm{rad}}=\sum_{\mathbf{k},\lambda}\hbar\omega_{k}a_{\mathbf{k}\lambda}^{\dagger}a_{\mathbf{k}\lambda}$
and $H_{\mathrm{int}}=-e\mathbf{A\cdot p}/m_{0}c=H_{\mathrm{em}}+H.c.$ is the optical interaction term, which is
linear in both the vector potential $\mathbf{A}$ and the electron momentum $\mathbf{p}$ and can be decomposed
into a photon emission term $H_{\mathrm{em}}$ and its Hermitian conjugate. For simplicity, we assume that the
dots $L$ and $R$ are identical, with cubic crystal structure and with aligned main crystal axes. We choose the
$z$ axis parallel to the quantum dot growth direction (e.g., [001]). If the quantum dot confinement is stronger
in the $z$ direction than in the $xy$ plane, $z$ defines the spin quantization axis and heavy-hole (hh) and
light-hole (lh) states are energetically split by $\Delta_{\mathrm{hh-lh}}$ (typically
$\Delta_{\mathrm{hh-lh}}\sim 10\, \mathrm{meV}$). We consider a hh ground state, with angular momentum
projection $\pm 3/2$ in terms of electron quantum numbers. We further focus on the strong-confinement regime,
where the dot radius is smaller than the  exciton Bohr radius.

The quantum dots in both spin-LEDs are prepared in a state $\ket{\chi_{\alpha}}$, where two excess holes occupy
the lowest hh level in each dot. This initial state, which can be generated by applying an appropriate bias
voltage across the LED, has several advantages. Firstly, electrons with arbitrary spin states can recombine
optically, as demonstrated for electron spin detection in a recent experiment.~\cite{guendogdu:04} Secondly, the
$z$ component of the total hole spin vanishes. This is a consequence of the fact that in quantum dots the hh-lh
exciton mixing due to the electron-hole exchange interaction $\Delta_{\mathrm{ehx}}$ is determined by a small
parameter $\Delta_{\mathrm{ehx}}/\Delta_{\mathrm{hh-lh}}\sim 0.01$. Thus, injected spin-polarized electrons give
rise to circularly polarized $X^+$ luminescence. This remains true for dots with asymmetric confinement in the
$xy$ plane, in stark contrast to the case with an electron and only one hole in the dot,~\cite{takagahara:00}
where the good exciton eigenstates are horizontally polarized and are split in energy typically by
$\delta_{\mathrm{ehx}}\sim 0.1\,\mathrm{meV}$. Thus, the electron-hole exchange interaction can be neutralized
by initially providing {\it two} holes. Interband mixing  (e.g., hh and lh states) in strongly anisotropic dots
reduces the maximum circular polarization of photons emitted from spin-polarized electrons \cite{pryor:03} and
reduces the fidelity of our scheme. However, because the interband transition probability for lh states is three
times smaller than that for hh states, and hh-lh mixing is typically controlled by some small para\-meter in
slightly elliptical dots,~\cite{takagahara:00} we neglect lh transitions.

\subsection{Electron injection and photon emission}
We first describe the dynamics of the electron injection and recombination in the two dots using a master
equation. The rate for the injection  and the subsequent relaxation of electrons into the conduction band ground
state in the dot $\alpha$ is denoted by $W_{e\alpha}$. It has been demonstrated that this entire process is spin
conserving and occurs much faster than the optical recombination~\cite{seufert,guendogdu:04}, which is described
by the rates $W_{p\alpha}$. Typically, $W_{p\alpha}\sim 1\:(\mathrm{ns})^{-1}$ and $W_{e\alpha}\sim
0.1\:(\mathrm{ps})^{-1}$ for the incoherent transition rates. We solve the master equation for the classical
occupation probabilities and obtain the probability that two photons are emitted after the injection of two
electrons  into the dots  at $t=0$,
\begin{equation}
P_{2p} =  \prod_{\alpha = L,R}\frac{W_{e\alpha}(1-e^{-tW_{p\alpha}})-
W_{p\alpha}(1-e^{-tW_{e\alpha}})}{W_{e\alpha}-W_{p\alpha}}.
\end{equation}
For $W_{p\alpha} \ll W_{e\alpha}$, $P_{2p}\approx \prod_{\alpha = L,R}(1-e^{-tW_{p\alpha}})$. After photon
emission, bipartite photon entanglement is achieved by a measurement of the hole spins  as we describe below and
the initial state is finally restored by injection of two holes into each of the two dots. We estimate the
production rate of entangled photons in a setup to test some of the proposed electron
entanglers.\cite{andreev:01,lesovik:01,recher:02,bena:02,bouchiat:03,saraga:03,recher:03,saraga2:04} For
example, electron spin singlets $|\Psi^-\rangle =(\spupdown -\spdownup)/\sqrt{2}$ are produced by the Andreev
entangler~\cite{andreev:01} with an average time separation $\Delta t \sim 10^{-5}\mathrm{s}$, while for the
entangler based on three quantum dots,~\cite{saraga:03} $\Delta t \sim 10^{-8}\mathrm{s}$. The two  electrons of
a singlet typically are injected into the current leads with a relative time delay $\tau \simeq
10^{-13}\mathrm{s}$ for both of these entanglers. Because $\tau, W_{p\alpha}^{-1} \ll \Delta t$, photons
originating from a single pair of entangled electrons can be identified with high reliability. In the steady
state, the generation rate of entangled photons is determined by the rate at which entangled electron pairs
leave the entangler, $1/\Delta t$.

\subsection{Electron spin dynamics}
Relaxation and decoherence is taken into account for the two spins by the single-spin Bloch
equation.~\cite{burkard:2003} Given that the electrons are in different leads, they interact with different
environments (during times $t$ and $t'$, respectively). Therefore, we consider different magnetic fields
$\mathbf{h}$ and $\mathbf{h'}$, enclosing an angle $\beta$, each acting on an individual spin. We calculate the
two-spin density matrix $\chi(t,t')$ and obtain for the singlet fidelity
$f=4\langle\Psi^-\left|\chi(t,t')\right|\Psi^-\rangle$ (given in Ref.~\cite{burkard:2003} for $t=t'$ and $\beta
=0$),
\begin{eqnarray}\nonumber
f & = & 1-\mbox{cos}\beta\, a a'P P'+
e_1\left[e'_2\mbox{sin}^2\beta\,\mbox{cos}(h't')
+e'_1\mbox{cos}^2\beta \right]\\ \nonumber
 & &  + e_2e'_1\mbox{sin}^2\beta\,\mbox{cos}(ht)
+ e_2e'_2\left[2\,\mbox{cos}\beta\,\mbox{sin}(ht)\,\mbox{sin}(h't')
 \right.\\
 & & + \left.\left(\mbox{cos}^2\beta\,+1\right)\mbox{cos}(ht)\,\mbox{cos}(h't')
\right] ,
\end{eqnarray}
where for the first (second) spin $e_i=e^{-t/T_i}$ ($e'_i=e^{-t'/T'_i}$), $a=1-e_1$ ( $a'=1-e'_1$), $P$ ($P'$)
is the equilibrium polarization, and $T_2$  and $T_1$ ($T'_2$ and $T'_1$) are the spin decoherence and
relaxation times, respectively. For $t \ll T_1,T_2$ and $t' \ll T'_1,T'_2$ (in bulk GaAs $T_2\sim
100\,\mathrm{ns}$ has been measured \cite{kikkawa:1998a} and, typically, $T_1\gg T_2$), the electrons form a
nonlocal spin-entangled state after their injection  into the dots $L$ and $R$ and after their subsequent
relaxation to the single-electron orbital ground states $\phi_{c\alpha}(\mathbf{r}_{c\alpha},\sigma)$. A local
rotation of one of the two spins in the leads (for $\mathbf{h} \neq \mathbf{h}'$) enables a transformation of
$|\Psi^-\rangle$ into another (maximally entangled) Bell state $|\Psi^+\rangle =(\spupdown +
\spdownup)/\sqrt{2}$ or $\ket{\Phi^{\pm}}=(\spupup\pm\spdowndown)/\sqrt{2}$. This can be achieved, e.g., by
controlling the local Rashba spin-orbit interaction in the current leads.~\cite{egues:2002,burkard:2003}

\section{Optical transitions \label{sec:Optic}}
The optical recombination processes of the two electrons occur
independently, except for the entanglement of the spin wave functions.
We consider one single branch $\alpha=L,R$ of the apparatus and omit
the index $\alpha$. The state of the single quantum dot which is
charged with two hhs in the orbital ground state and into which
a single electron with spin $\sigma$ has been injected is given by
\begin{equation}
\ket{e,\sigma}  =  \int\mathrm{d}^{3}r_{c}
\phi_{c}^{*}(\mathbf{r}_{c},\sigma) b_{c\sigma}^{\dagger}(\mathbf{r}_{c})\ket{\chi}.
\label{eq:exstate}
\end{equation}
Here, $b_{c\sigma}^{\dagger}(\mathbf{r}_{c})$  creates an electron with
spin $S_{z}=\sigma/2=\pm1/2$
at $\mathbf{r}_c$ in the ground state of the dot,
$\ket{\chi}=\sum_{\tau\neq\tau'}\int \mathrm{d}^{3}r_{v1} \mathrm{d}^{3}r_{v2}\phi_{v}(\mathbf{r}_{v1},\tau;\mathbf{r}_{v2},\tau') b_{v\tau}(\mathbf{r}_{v1})b_{v\tau'}(\mathbf{r}_{v2})\ket{g}$,
where $\ket{g}$ is the electrostatically neutral ground state of the
quantum dot, and $\phi_{v}(\mathbf{r}_{v1},\tau;\mathbf{r}_{v2},\tau')$
is the orbital part of the two-hole wave function. In the strong-confinement
regime where Coulomb correlations are negligible, $\phi_{v}$ is a product
of the single-particle valence band states. The labels $\tau,\,\tau'$
denote the hh spin component $S_{z} = \tau /2 = \pm1/2$ that factor out
for angular momentum $J_{z}=\pm 3/2$. We now calculate the emission matrix
element $\bra{f}H_{\mathrm{em}}\ket{i}$ with initial state
$\ket{i}=\ket{e,\sigma}\otimes\ket{\dots,n_{\mathbf{k}\lambda},\dots}$
and final state $\ket{f}=b_{v\tau'}(\mathbf{r}_{v2})\ket{g}\otimes\ket{\dots,n_{\mathbf{k}\lambda}+1,\dots}$,
where $\ket{\dots,n_{\mathbf{k}\lambda},\dots}$ is a Fock state of the
electromagnetic field, typically the photon vacuum. Because of quantum
mechanical selection rules, the optical transitions connect only states
with the same spin such that $\tau'\neq\sigma$.
In the envelope-function and dipole approximations,~\cite{biexcitons}
\begin{equation}
|\bra{f}H_{\mathrm{em}}\ket{i}|  =
\frac{e}{m_{0}c}\, A_{0}(\omega_{k})\sqrt{n_{\mathbf{k}\lambda}+1}\,
\left|\mathbf{e}_{\mathbf{k}\lambda}^{*}\cdot\mathbf{p}_{cv}^{*} C_{eh}\right|,
\label{eq:emmatrixelement}
\end{equation}
where $\mathbf{p}_{cv}^{*}=\mathbf{p}_{vc}$ is the inter-band momentum
matrix element,
$\mathbf{e}_{\mathbf{k}\lambda}$ is the unit polarization vector with
$\lambda=\pm1$ for circular polarization $|\sigma_{\pm}\rangle$,
$A_{0}(\omega_{k})=(\hbar/2\epsilon\epsilon_{0}\omega_{k}V)^{1/2}$,
and $C_{eh}=\int\mathrm{d}^{3}r\,\psi_{c}^{*}(\mathbf{r},\sigma)\psi_{v}(\mathbf{r},\sigma)$,
where $\psi_{n}$ is the envelope function of a carrier in the band $n=c,v$.
For cubic symmetry, $\mathbf{e}^{*}_{\mathbf{k}\lambda}\cdot\mathbf{p}_{cv}^{*} = p_{cv}(\cos\theta-\sigma\lambda)e^{-i\sigma\phi}/2 \equiv  p_{cv}m_{\sigma\lambda}(\theta,\phi)$,
where $\theta$ and $\phi$ are the polar and the azimuthal angle of the
photon emission direction, respectively.
With the transition $\ket{e,\sigma}\rightarrow b_{v-\sigma}(\mathbf{r}_{v2})\ket{g}$, a photon
\begin{equation}
|\sigma,\theta,\phi\rangle=N(\theta)(m_{\sigma,+1}(\theta,\phi)|\sigma_{+}\rangle+m_{\sigma,-1}(\theta,\phi)|\sigma_{-}\rangle)
\label{eq:photonstate}
\end{equation}
is emitted into the direction $(\theta , \phi)$. Here, $N(\theta)=[2/(1+\cos^{2}\theta)]^{1/2}$ is a
normalization factor. Eq.~(\ref{eq:photonstate}) shows that for $\theta=0$, a spin-up ($\sigma=+1$) electron
generates a $|\sigma_{-}\rangle$ photon, whereas a $|\sigma_{+}\rangle$ photon is obtained from a spin-down
($\sigma=-1$) electron. The admixture of the opposite circular polarization increases with $\theta$, leading to
linear polarization for $\theta=\pi/2$. For $\theta\neq0$, the spin-inverted states $|+1,\theta,\phi\rangle$ and
$|-1,\theta,\phi\rangle$ have interchanged coefficients for $|\sigma_{+}\rangle$ and $|\sigma_{-}\rangle$, up to
a relative phase determined by the (global) phase factors $\exp{(-i\sigma\phi)}$. Note that in two-photon states
the azimuthal angles thus can provide a {\it relative} phase, as we exploit below.

\subsection{Entangled four-photon state}
The two photons produced at recombination are entangled with the two
holes which remain in the dots, due to the antisymmetric hole ground
state. By injecting a pair of electrons with spins polarized in the
$xy$ plane into the dots,\cite{twopairs}  a four-photon state of
the Greenberger-Horne-Zeilinger (GHZ) type \cite{peres:1998a} can be produced if
$T_{1,X}$ and $T_{2,X}$  exceed the exciton lifetime $\tau_X$.
For the two polarized electrons, only the electron spin orientation
in $z$ direction which satisfies the optical selection rules contributes
to the optical transition, respectively.
For circularly polarized photons emitted along $z$, the electron
Bell states give rise to the photon states
\begin{eqnarray}
\ket{\Psi^{\pm}} & \rightarrow & |\sigma_+ \sigma_- \sigma_- \sigma_+ \rangle
\pm |\sigma_- \sigma_+ \sigma_+ \sigma_- \rangle,\label{eq:ghz1}\\
\ket{\Phi^{\pm}} & \rightarrow & |\sigma_- \sigma_- \sigma_+ \sigma_+ \rangle
\pm |\sigma_+ \sigma_+ \sigma_- \sigma_- \rangle,\label{eq:ghz2}
\end{eqnarray}
where the first two entries indicate the first photon pair (L,R) and the third and fourth entry the second
photon pair (L,R), respectively. Normalization has been omitted for simplicity. Yet, the second photon pair is
generated by neutral excitons and is thus exposed to the same problems as the biexciton decay cascade in
asymmetric quantum dots. Here, a cavity can be used to maintain the GHZ state since the energy entanglement of
the second photon pair can be erased,~\cite{stace:2003a} and $\tau_X$  can be shortened due to the Purcell
effect to reduce exciton polarization decoherence.

\subsection{Entangled two-photon state}
\begin{figure}[t]
\centerline{\includegraphics[width=8cm]{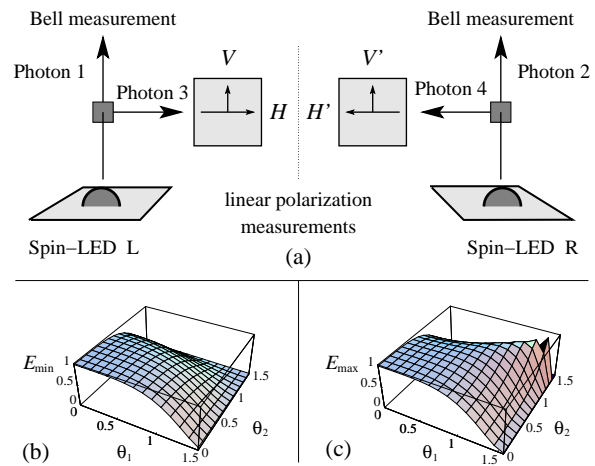} \vspace{1mm}} \caption{(Color online) (a)
Schematic setup to obtain bipartite entanglement of photons 1 and 2 by measuring the photons 3 and 4 of the GHZ
state in bases of linear polarizations $H,V$ and $H',V'$, respectively (see the text). In (b) and (c), we show
the von Neumann entropy (b) $E = E_{\mathrm{min}}$ and (c) $E = E_{\mathrm{max}}$ as a function of the polar
angles $\theta_1$ and $\theta_2$ for photon emission.
 $E$ oscillates
between (b) and (c) as a function of $\phi_1$ and $\phi_2$, as explained in the text. The photon-polarization
entanglement is maximal for $\theta_1 = \theta_2 = 0$, whereas  for $\theta_i = \pi /2$ entanglement is absent.
In (c), $ E_{\mathrm{max}}=1$ for the continuous set of directions $\theta_{1}=\theta_{2}\in[0,\pi/2)$.
}
\label{fig:entropy}
\end{figure}

Full {\it bipartite} photon entanglement of the first photon pair is obtained, e.g., by directing the second
photon pair via secondary optical paths to a linear polarization measurement which is performed {\it before} the
first photon pair is measured,~\cite{imamoglupc} see Fig.~\ref{fig:entropy} (a). Even different bases
$\{|H\rangle ,\, |V\rangle\}$ and $\{|H'\rangle ,\, |V'\rangle\}$ can be chosen for the two photons of the
second pair. Note that the electron-hole exchange interaction in elliptical dots assists this projection into
linearly polarized eigenstates (along the major and the minor axis of the dots, respectively) already during the
lifetime of the remaining two excitons. While the loss of (linear) polarization coherence is tolerable for these
excitons, $T_{1,X}>\tau_X$ is required for entanglement of the first photon pair. This suggests that the scheme
presented here can be realized with typical quantum dots, see Ref.~\cite{tsitsishvili:2003a} and references
therein.

If the second photon pair is measured in the state $|HH'\rangle$ or $|VV'\rangle$, the electron Bell states have
given rise to the two-photon states
\begin{eqnarray}
\ket{\Psi^{\pm}} & \rightarrow & |\!+\!\!1,\theta_{1},\phi_{1}\rangle_{L}\,|\!-\!\!1,\theta_{2},\phi_{2}\rangle_{R} \nonumber\\
& & \pm\,|\!-\!\!1,\theta_{1},\phi_{1}\rangle_{L}\,|\!+\!\!1,\theta_{2},\phi_{2}\rangle_{R},\label{eq:2photonstate1}\\
\ket{\Phi^{\pm}} & \rightarrow & |\!+\!\!1,\theta_{1},\phi_{1}\rangle_{L}\,|\!+\!\!1,\theta_{2},\phi_{2}\rangle_{R} \nonumber\\
& &
\pm\,|\!-\!\!1,\theta_{1},\phi_{1}\rangle_{L}\,|\!-\!\!1,\theta_{2},\phi_{2}\rangle_{R}.\label{eq:2photonstate2}
\end{eqnarray}
Here, normalization has been omitted for simplicity. If the second photon pair is measured as $|HV'\rangle$ or
$|VH'\rangle$, $\pm$ is replaced by $\mp$ on the right-hand side of Eqs.\ (\ref{eq:2photonstate1}) and
(\ref{eq:2photonstate2}).

Obviously, above  two-photon states (\ref{eq:2photonstate1}) and (\ref{eq:2photonstate2}) are maximally
entangled for $\theta_{1}=\theta_{2}=0$. For $\theta_{1}=\theta_{2}\in(0,\pi/2)$, the total relative phase
factor between the two-photon states in Eq.~(\ref{eq:2photonstate1}) is $\exp(i\gamma+2i\Delta\phi)$. Here,
$\Delta\phi=\phi_{1}-\phi_{2}$,  and the relative phase of the two-electron states is $\gamma=\pi$ for
$\ket{\Psi^{-}}$ and $\gamma=0$ for $\ket{\Psi^{+}}$. For Eq.~(\ref{eq:2photonstate2}), the relative phase
factor is $\exp[i\gamma+2i(\phi_{1}+\phi_{2})]$, with $\gamma=\pi$ for $\ket{\Phi^{-}}$ and $\gamma=0$ for
$\ket{\Phi^{+}}$. By tuning the relative phase factors in Eqs.\ (\ref{eq:2photonstate1}) and
(\ref{eq:2photonstate2}) to $-1$, two circularly polarized photons can be recovered for $\theta_1=\theta_2 \in
(0,\pi/2)$ from the elliptically polarized single-photon states due to quantum mechanical
interference.\cite{ghzent} Thus, maximal entanglement is transferred from two electron spins to the
polarizations of two photons for certain ideal emission angles. For $\ket{\Psi^{-}}$ ($\ket{\Psi^{+}}$),
$\Delta\phi=0$ ($\Delta\phi=\pi/2$) needs to be satisfied $\mathrm{mod}\pi$, whereas  the condition for
$\ket{\Phi^{-}}$ ($\ket{\Phi^{+}}$) is $\phi_{1}+\phi_{2}=0$ ($\phi_{1}+\phi_{2}=\pi/2$)  $\mathrm{mod}\pi$. For
$\theta_1=\theta_2 = \pi /2$ these two-photon states vanish completely due to destructive interference.

\section{Photon entanglement as a function of emission directions \label{sec:Entanglement}}
For arbitrary emission directions of the two photons, the degree of
polarization entanglement can be quantified by the von Neumann entropy
$E=-\text{tr}_{2}(\tilde{\rho}\log_{2}\tilde{\rho})$.
Here, $\tilde{\rho}=\text{tr}_1\rho$ is the reduced density matrix of the two-photon state $\rho$
with the trace $\text{tr}_1$ taken over photon 1. For a maximally entangled two-photon state $E=1$,
while $E=0$ represents a pure state
$\tilde{\rho}$ (which implies the absence of bipartite entanglement).
If the two electrons recombine after times much shorter than the spin
lifetimes $T_1,\,T'_1,\,T_2,\,T'_2$, $E$ oscillates for
Eq.~(\ref{eq:2photonstate1}) as a function of $\Delta\phi$ of the
two emitted photons between a minimal value,
\begin{eqnarray}
 E_{\mathrm{min}} &=& \log_{2}(1+x_{1}x_{2})
- \frac{x_{1}x_{2}\log_{2}(x_{1}x_{2})}{1+x_{1}x_{2}},
\end{eqnarray}
%$E_{\mathrm{min}}$ (given in Ref.~\cite{biexcitons}),
and a maximal value,
\begin{equation}
E_{\mathrm{max}}=\log_{2}(x_{1}+x_{2})-\frac{x_{1}\log_{2}(x_{1})}{x_{1}+x_{2}}-\frac{x_{2}\log_{2}(x_{2})}{x_{1}+x_{2}},\label{eq:emax}
\end{equation}
where $x_{i}=\mbox{cos}^2\theta_{i}$, which is (only) obtained for the ideal angles $\phi_1$ and $\phi_2$
mentioned above; see Fig.~\ref{fig:entropy} (b) and (c). For Eq.~(\ref{eq:2photonstate2}), $E$ oscillates
between $E_{\mathrm{min}}$ and $E_{\mathrm{max}}$ as a function of $\phi_1 + \phi_2$. As expected,
$E_{\mathrm{max}}=1$ for all $\theta_{1}=\theta_{2}\in[0,\pi/2)$. The discontinuity in  $E_{\mathrm{max}}$ for
$\theta_1=\theta_2=\pi /2$ is due to the vanishing two-photon state.

\section{Conclusions \label{sec:Concl}}
We have studied the transfer of entanglement from electron spins to photon polarizations. We have discussed the
generation of entangled four-photon and two-photon states via the injection of spin-entangled electrons into
quantum dots charged with two excess holes. We have proposed a scheme to achieve complete entanglement transfer
from two electron spins to two photons. We have shown that this scheme can even be realized with quantum dots
exhibiting an exciton exchange splitting. We have shown the dependence of the photon entanglement on the
emission angles and identified the conditions for maximal entanglement. This offers the possibility to
efficiently test Bell's inequalities for electron spins. In addition, our results show that a continuous set of
directions exist along which entanglement is maximal. Finally, similar schemes to produce entangled photons can
be realized using two tunnel-coupled dots~\cite{gywat} instead of two isolated dots. In such a setup, it is
essential that tunnel coupling is provided for the conduction-band electrons, whereas the valence-band holes are
not tunnel coupled and thus localized in the individual dots. After a positively charged exciton is created in
each of  the two dots, the spin entanglement is provided from the singlet ground state of the delocalized
electrons and can be transferred to the photons, similarly as described in this work.

We thank  A. Imamo\=glu, G. Burkard, F. Meier,  P. Recher,  D. S. Saraga, V. N. Golovach, and D. V. Bulaev for
discussions. We acknowledge support from DARPA, ARO, ONR, NCCR Nanoscience, and the Swiss NSF.


\begin{thebibliography}{99}
\bibitem{ghzent}Such ideal angles can analogously be found for the
four-photon GHZ states.
\end{thebibliography}
\end{document}